# The role of cosmic rays in the Earth's atmospheric processes

DEVENDRAA SIINGH[1,2*] and R P SINGH[3]

[1]Indian Institute of Tropical Meteorology, Pune-411 008, India
[2]Center for Sun-Climate Research, Danish National Space Institute, Copenhagen, Denmark.
[3]Vice-Chancellor, Veer Kunwar Singh University, Ara-802301 (Bihar), India
**\* e-mail,** devendraasiingh@tropmet.res.in**;** devendraasiingh@gmail.com

## Abstract

In this paper, we have provided an overview of cosmic ray effects on terrestrial processes such as electrical properties, global electric circuit, lightning, cloud formation, cloud coverage, atmospheric temperature, space weather phenomena, climate, etc. It is suggested that cosmic rays control short term and long term variation in climate. There are many basic phenomena which need further study and require new and long term data set. Some of these have been pointed out.

## 1. Introduction

The Sun is the chief driving force of the terrestrial atmospheric processes. Hence, any variation in atmospheric processes is attributed to variation in solar radiation and its modulation by the Earth's orbital motion. However, the observed variations cannot be explained fully by the variation in solar radiation. Some of these variations have been attributed to cosmic rays of both galactic and solar origin. It is interesting to note that solar energy flux reaching the Earth's orbit is Fs = $1.36 \times 10^3$ Wm$^{-2}$ whereas the cosmic ray energy flux (particles with energy $\geq$ 0.1 GeV) is $F_{CR}$ = $10^{-5}$ Wm$^{-2}$ [1]. Thus, energy input by cosmic rays in the Earth's atmosphere is about $10^{-9}$ times to that of solar energy and hence it is unlikely that cosmic rays could influence the atmospheric processes. However, cosmic rays are the only source of ion production in the lower atmosphere, which is confirmed from the measurements of *Ermakov and Komozokov* [2]. Therefore, the processes depending on the electrical properties of the atmosphere such as atmospheric electric current, lightning production, cloud and thundercloud formation, etc can be affected by cosmic rays.

The low energy cosmic ray particles (energy $\leq$ 15 GeV) undergo a 11 year modulation and the flux of cosmic ray particles with energy in the range 0.1 – 15 GeV decreases more than two folds when activity period changes from the minimum to the maximum [3]. About 95% of cosmic rays particles fall in this energy range which contain more than 60% of all cosmic ray particle energy [3,4]. As the total energy is used in exciting and ionizing air atoms, it is expected that the effect of cosmic rays should be more dominant during the period of minimum activity.

Observational evidences show that the total cloudiness and precipitation reduced when cosmic rays fluxes in the interplanetary space and the atmosphere decreased (Forbush decrease) [5-7]. Further, diverse reconstructions of past climate change revealed clear associations with cosmic ray variations recorded in cosmogenic isotope archives [*Kirkby,* 4, *and references therein*]. For example, the decade 1690-1700 was the coldest during the last 1000 years and during this period $^{10}$Be concentration had the largest peak. The variation in $^{10}$Be is a signature of changes in cosmic ray flux. Global temperature and $^{10}$Be concentration have opposite trends. Sevensmark [8] plotted magnitude of change in $^{10}$Be concentration and change in temperature during the period of Maunder minimum and showed a striking similarity. The low solar magnetic activity during the Maunder minima and earlier periods might have been among the principal cause of the Little Ice Age [4,9]. During the Maunder minima, the absence of strong magnetic field region on the surface of the Sun might have affected the solar wind flow and hence modified the characteristic features of cosmic rays incident on the Earth's atmosphere. *Kirkby* [4] had presented an association of high GCR flux with cooler climate, and low GCR flux with a warmer climate.

Ionization produced by cosmic rays in the troposphere and stratosphere produces ultra-fine aerosols which may act as cloud condensation nuclei [10-13]. The aerosol layers have also significant effect on the Earth's atmosphere heat balance through scattering of solar beam in the forward direction and hence effective reduction in solar constant [14]. The short life time (~ few days) of aerosols in the troposphere results in significant spatial and temporal variations in aerosol particle concentration, size and composition. The high variability leads to one of the largest uncertainties of anthropogenic climate forcing. The variation in aerosols in the lowest few kilometers of the atmosphere leads to local turbulent fluctuations of space charge density which impose a time varying electric field. This electric field at times may be comparable in magnitude to the electric field maintained by global thunderstorm activity and thus affects the global electric circuit [15-18].

In this paper, we have reviewed the present status of the role of cosmic rays in the Earth's atmospheric processes. Section two briefly discusses role of cosmic rays on the Earth's climate. We briefly describe possible physical processes caused by cosmic rays, correlation in cloud cover variation and cosmic ray activity. In the third section, cosmic rays and ion production are described, whereas discussion on cosmic rays and aerosols are given in the fourth section. Lightning is another process which affects climate and electrical properties of the atmosphere; hence it is briefly described in section five. Ozone distribution directly influences climate, so role of cosmic rays on ozone distribution is given in section six, whereas role of CR on space-weather studies is discussed in section seven. Summary of present study is given in the last section.

## 2. Cosmic ray and the Earth's climate

The total solar irradiance reaching the Earth's atmosphere is the main driving agent for the variation in the Earth's climate. There are three probable mechanisms which are thought to link solar variability with climate [3]. These are (a) changes in total solar irradiance leading to changes in heat input to the lower atmosphere, (b) solar ultraviolet radiation coupled to change in ozone concentration heating the stratosphere, and (c) galactic cosmic rays. These effects are modulated by long term solar magnetic activity, by changes of the source of galactic cosmic rays and by changes in the Earth's magnetic field [19-21].

The effect of cosmic rays on climate could be in three ways: (a) through changes in the concentration of cloud condensation nuclei, (b) thunderstorm electrification, and (c) ice formation in cyclones. The concentration of cloud condensation nuclei depends on the light ions produced during cosmic ray ionization [3,22]. *Svensmark et al.* [23] based on a laboratory experiment in which a gas mixture equivalent to chemical composition of the lower atmosphere was subjected to UV light and cosmic rays, reported that the released electrons promoted fast and efficient formation of ultra-fine aerosol particles which may grow to become cloud condensation nuclei. Air-craft based ion mass spectrometer measurements in the upper troposphere have also shown an evidence for cosmic ray induced aerosol formation [24,25].

### 2.1. Cosmic ray influence on global temperature

It has been suggested that the global warming may result in enhanced convective activity of thunderstorm and in turn increased thunderstorm production on a global scale [26]. The temperature during the last century on the Earth's surface increased by ~ $0.6^{o}$ C [27], however its effect on lightning is not yet quantified. *Stozhkov* [28] had suggested two mechanisms for heating namely solar influence on the weather and climate, and the influence of human activities on atmospheric processes (such as greenhouse effect). The physical mechanism of the greenhouse gases has been understood whereas the mechanism of solar influence on weather and climate requires detail study.

*Lean et al.* [29] have calculated the correlation between Northern hemisphere surface temperature and solar irradiance (reconstructed from solar indices) from 1610 to 1994 and showed the coefficient to be 0.86 in the pre-industrial period from 1610 to 1800. Extending this correlation, they have suggested that solar forcing may have contributed about half of the observed $0.6^0C$ surface warming since 1860 and one third of the warming since 1970. The remaining change in global temperature during the pre-industrial period may be due to cosmic rays forcing and other natural processes. *Svensmark* [8] studied the variation of 11 year average temperature in terms of percent change in cosmic ray decrease and sunspot number during the period 1935-1995 and showed that the best correspondence between solar activity and temperature seems to be between solar cycle length and the variation in cosmic ray flux. Further, the closest match with ion chamber cosmic ray data suggests that the high energy cosmic rays responsible for ionization in the lower atmosphere play major role and hence it is suggested that this part of the atmosphere should be looked into for a physical effect.

Galactic and solar cosmic rays influence physical-chemical process (reactions) in the lower atmosphere including cloudiness density changes and atmosphere cloud coverage and thus control the variability of atmosphere transparency and thereby affect solar radiation flax reaching in the lower atmosphere. Clouds reflect both the incoming solar radiation flux upward and the Earth's thermal radiation back to it, thereby control thermal energy input in the lower atmosphere. Thus establishing a link between cosmic rays and temperature.

**2.2. Cosmic rays - cloud variability and involved mechanisms**

The cosmic rays and low altitude cloud (<3 km altitude) variation derived from satellite measurements were reported to be correlated around the cosmic rays minimum of 1990 [30,31]. The limitations of the data analysis and acquisitions have been the subject of considerable discussions [32-35]. *Kernthaler et al.* [32] extended the approach of *Svensmark and Friis-Christensen* [30] to the stable data set during the period 1985-1989 for different cloud types but could not find any clear relationship between individual cloud types and cosmic ray flux. Further, the amplitude of apparent correlation decreased when high latitude data set is included in the analysis although ionization by cosmic rays is high at high latitudes. *Sun and Bradley* [34] re-examined the surface-based cloud data and newer satellite data for an extended period but they also could not find any meaningful relationship between cosmic ray intensity and cloud cover over tropical and extra-tropical land areas.

*Kristjansson and Kristiansen* [36] have analyzed data set of the ISCCP - D2 satellite recorded during 1989 to 1993, and found little statistical evidence of a relationship between GCRs and low-altitude-marine-cloud cover in mid latitudes. *Harrison and Stephenson* [37] showed the existence of cosmic ray effects on clouds on long time scales with less variability than the considerable variability of daily cloudiness.

*Kuang et al.* [38] computed the variation of cloud optical thickness and their relationship with cosmic ray intensity and El Nino-Southern Oscillation (ENSO) activity over the period 1983 -1991 and showed that the variation in cloud optical thickness could easily be explained by the E1 Nino activity rather than variations in cosmic ray intensity. *Farrar* [39] analyzed ISCCP- C2 data for the period July 1986 to June 1991 and showed that the global and regional cloud cover variations were associated with the El-Nino of 1986-1987 and were particularly apparent in the Pacific region.

Numerical modeling and satellite observations suggested that a 1% change in the total composition of the Earth's cloud cover corresponds to 0.5 W/m$^2$ changes in net radiative forcing [40]. Therefore, a change in global cloudiness by approximately 3.0 % [31] may correspond to 1.5 W/m$^2$ radiative forcing during the year 1987 - 1990. During this period cosmic rays changed by ~ 3.5 % [31]. *Rossow and Cairns* [40] have also calculated the approximate radiative forcing by taking into account the running mean of 11- years average increase of cosmic rays from 1975 - 1989 in between 0.6 - 1.2 %, which comes out to be 0.3-0.5 W/m$^2$ change in cloud forcing. The direct influence of changes in solar irradiance is estimated to be only $0.1^0$ C [29]. The cloud forcing from the above data comes out to be about 0.2-0.5$^0$ C. In this calculation it is assumed that the whole cloud volume is affected by solar activity. In real situation this may not be valid. The above discussion shows that an increase in the cloud cover may result in lower temperatures.

*Solanki* [41] and *Le Mouel et al*. [42] claimed a good correlation between geomagnetic field changes, solar irradiance and global temperature. *Vieira and da Silva* [43] reported a cooling effect of ~ 18Wm$^{-2}$ in the inner region of Southern

Hemisphere Magnetic Anomaly (SHMA) and a heating effect of same magnitude in the outer region of SHMA. They also reported that correlation between net radiative flux and GCR increases in the inner region of SHMA. The presence of SHMA involves stronger cosmic ray / cloud interaction in the lower field region and enhances cooling, whereas in the outer region weaker cosmic rays / cloud interaction leads to less cooling or heating effect. Recently, *Courtillot et al.* [21] showed a connection between climate and geomagnetic field variations at various time scales (secular variation ~ 10 - 100 years, historical and archeomagnetic changes ~ 100 – 5000 years, and excursion and reversals ~ $10^3$ - $10^6$ years). However involved mechanism is not known.

Despite these criticisms, cosmic rays - climate relation is finding better acceptance day by day. This may be due to emerging physico-chemical mechanisms. *Yu* [44] has explained differential cosmic rays effects on low and high clouds in terms of ion-induced particle production and its recombination rates which are proportional to their concentrations. The ion life time is greater at lower altitude where the cosmic ray produced ionization rate is smaller. In the lower atmosphere aerosol particle density is larger and hence cloud condensation nuclei would be formed easily.

There are two mechanisms which link cosmic rays with cloud. In the first mechanism, cosmic ray produced ions influence the production of new aerosol particles in the troposphere, which may grow and eventually increase the number of cloud condensation nuclei. These nuclei act as seed for the cloud droplet formation. In the second mechanism, cosmic ray produced ionization modulates the global electric current which influences cloud properties through charge effects at cloud surface on droplet freezing and other microphysical processes.

Cosmic rays produced ions in the atmosphere may enhance the birth and growth of aerosol particles by a mechanism known as ion-induced nucleation mechanism [3]. In fact, newly produced aerosols consist of small clusters containing as few as two molecules, which may grow by condensation of trace vapours and water to a critical size or may evaporate before reaching the critical size. Once critical size is reached, the cluster is more likely to grow by further condensation; below the critical size, it is more likely to shrink by evaporation [4]. The presence of charge stabilizes the cluster through Coulomb attraction, and reduces the critical size. This process is known as ion-induced nucleation [10,45-47].

The presence of charge accelerates the early growth process due to enhanced collision probability and thereby increases the fraction of particles that survive removal by coagulation before reaching the critical size. Thus, charge plays significant role by enhancing the survival of new particles to CCN sizes. However, only a small fraction of new particles could reach the minimum size to be effective CCN and rest are lost by scavenging on existing aerosols. *Tripathi and Harrison* [12,48] have shown that electro-scavenging effect increases as the aerosol radius decreases below one micron and the effect is independent of sign of the aerosol charge, because electrical image force dominates at the nearby distances from the droplet [11].

Model calculations [10,45,49] showed that a 20% variation in the ionization rate in the lower atmosphere could lead to a change in concentration of 3 to 10 nm diameter aerosols of about 5 to 10% and some of them may contribute to the CCN population. However, the fraction of CCN originating from CR ionization will depend upon many factors including availability of condensable gas, direct source of CCN and cloud processing.

The presence of cloud layer in the atmosphere causes perturbation in the electric field and vertical current, which are global electric circuit parameters. These perturbations cause the upper part of the cloud to become more positively charged than the clear air above it, with a gradual return to quiescent values about 200 m above the cloud [3]. Due to removal of small ions from within the clouds, a large conductivity gradient develops, which ultimately results in the space charges accumulation at the cloud-air boundaries. Space charges form quite larger (~ 100 electronic charges) equilibrium droplet charges at cloud boundaries [50]. This is called as ion-aerosol near-cloud mechanism [3,51]. The space charge and low conductivity environment around the cloud prevent the rapid neutralization of such droplets. The aerosol charging in this mechanism is coupled with GEC changes and hence the aerosol-cloud-microphysics at long distances away from the source disturbance could also be affected.

In the presence of high conductivity gradients near the clouds, the vertical conduction current generates highly charged droplet. When these droplets evaporate they leave behind highly charged and coated `evaporation nuclei' [*Kirkby,* 4, *and references therein*]. Some of these become ice nuclei and induce ice formation. The cloud-tops frequently contain a great deal of super cooled liquid water (~ - 40$^o$C). Charged aerosol ice nuclei will induce freezing [52]. Thus, an increased GCR intensity may lead to increased ice particle formation in clouds.

In order to understand the role of cosmic rays in cloud formation, ice particle formation and freezing mechanism of polar cloud, an experiment called CLOUD (Cosmic Leaving Outdoor Droplet) is being set up at the biggest particle-physics laboratory in the world CERN in Geneva, Switzerland. A beam of particles from CERN's Proton Synchrotron will be used to mimic cosmic rays. Particle source is adjustable to simulate GCRs at any altitude and latitude. The chamber has various control systems so as to simulate atmospheric conditions in temperature, pressure, gas/vapour constituents available at any altitude, latitude or longitude. Field cage is used to generate electric field identical to fair-weather electric fields of the atmosphere. During the experiment, attempts will be made to measure the effect of ionizing particles on the formation rate of ultra-fine condensation nuclei and after introducing trace gases ($H_2SO_4$, $HNO_3$, $NH_3$, etc) nucleation rate will be measured as a function of different parameters available in the atmosphere. Then the effect of cosmic rays on the growth of CN into CCN i.e. from ~ 5nm diameter to ~ 100 nm diameter will be measured. These experiments are to understand ion-induced nucleation of new aerosols and then activation of CCN into cloud droplets [4].

The other set of experiments would be to understand cloud microphysics such as ice particle formation, collision efficiencies of aerosols and droplets and freezing mechanism of polar stratospheric clouds. The success of these experiments will shed light on the role of cosmic rays on terrestrial climate and involved physical mechanisms.

**3. Cosmic rays, ion production and electrical properties of Earth's atmosphere**

The cosmic rays being high energy charged particles penetrate into the lower atmosphere and are also filtered by the geomagnetic field. The filtering effect becomes variable in time with the magnetospheric currents that grow during the periods of magnetic activity allowing particles of a given energy to penetrate to lower latitudes, where these particles produce ionization. *Stozhkov* [28] using the experimental data of ion concentrations evaluated cosmic ray fluxes at different altitudes and latitudes in the atmosphere for different geomagnetic cut off rigidities and concluded that the ion production rate linearly depends on air density and cosmic ray flux.

The ion production rate peaks at different heights depending on the energy of cosmic ray particle. It increases with latitude and decreases with solar activity. Below the 15 km altitude, the production is by cascades of secondary particles, particularly µ mesons [53]. In the polar stratosphere, the ion production is mainly by lower energy GCR (E in the range of 0·1–0·5 MeV), which are strongly modulated by the solar wind magnetic fields in the inter-planetary space as compared to 1–5 MeV component. The profiles of the ion production rate for cosmic rays for various geomagnetic latitudes during the solar cycle minimum using the data from *Nehar* [54] is presented in Figure 1. The ionization rate increases with geomagnetic latitudes, and both the height of the peak and the slope of the ionization rate after the peak also increase. Near the ground there is about 20% variation between the equatorial region and the higher latitudes. Ion mobility is lowered through water vapour nucleation on ions, followed by hygroscopic growth and through the ion attachment to coarse particles, namely those generated during volcanic eruption, combustion and dust re-suspension. The most abundant ion species observed in the troposphere and the stratosphere are complex cluster ions containing $H_2SO_4$, $H_2O$, $HNO_3$, $(CH_3)_2 CO$ and $CH_3CN$ molecules attached to core ions [55].

The cosmic rays are the major role players in the electrical properties of the atmosphere and the global electric circuit by manipulating atmospheric conductivity, ionospheric potential, vertical current and vertical ionospheric potential gradients. Decadal changes in the surface potential and ionospheric potential gradients correlated well with changes in GCR [56,57]. GCR modifies the atmospheric columnar resistance and hence the vertical current, which in clear weather is of the order of 1 pA/m$^2$. The additional charge pairs produced by cosmic rays enhance conductivity of the lower atmosphere and hence enhance the thunderstorm charging current by facilitating enhanced charge transfer. The vertical current flowing through the region of varying conductivity in the troposphere produces space charges $\rho(Z)$ which can be calculated using Gauss theorem $\nabla.E = \rho/\varepsilon_0$. Since $E_z = J_z/\sigma(z)$, the space charge $\rho(z) = \varepsilon_0 J_z d/dz(1/\sigma(z))$. Thus enhancement in cosmic rays intensity enhances production of space charges. Within the clouds, a large conductivity gradient exists, so layers of space charges develop at the top and bottom of clouds. The space charges near the cloud are in the form of charged aerosol particles. In the presence of electric fields associated with the development of space charges layers many charged aerosols accumulate and are attached to droplets. Thus, the perturbation in GEC also affects cloud formation.

The global circuit involving about 1400 A current and 400 MW power sustained by continuous thunderstorms links the upper atmosphere to the surface of the Earth and the changes in the electrical potential of the ionosphere are communicated directly by the air-earth current flowing through cloud layers to the Earth's surface [17,58]. Recently, *Rycroft et al.* [59] have discussed the global electric circuit model driven by thunderstorms and electrified shower clouds and have shown that each of them contribute almost equally in maintaining the ionospheric potential. They have presented the electric potentials and fields at different points in the circuit before, during and after cloud-to-ground lighting flashes of either negative or positive polarity, and also following a sprite. The physical consequence of this coupling for climate is not clear.

Figure (2) summarizes the range of processes that can result from the solar modulation of cosmic rays and link global atmospheric electrical circuit with global climate [57,58]. Cosmic rays and shower clouds cause ionization of atmospheric constituents and control the electrical potential and the vertical current flowing through cloud layers, which affect cloud microphysics leading to change in ice cloud and stratiform cloud and hence change in climate. Clouds act as reflector to the incoming radiations from the sun and escaping radiations from the Earth and hence also affect radiative transfer mechanism of the atmosphere leading to global temperature changes which control thunderstorm activity, and hence electrical system of the atmosphere, thus forming a closed system.

**4. Cosmic rays and aerosol**

Aerosol in the atmosphere are produced by several processes, some of them being gas-particle conversion, droplet particle conversion (evaporation of water droplets containing dissolved matter), bulk particles (smoke, dust or pollen, etc) from the surface carried by wind [60] and particles of various shapes and sizes carried upwards during volcanic eruption, bombings during war periods, etc. The nature and size of aerosol particles are height dependent and also have a significant effect on the heat balance of the Earth's atmosphere. The observation of size and density distribution of aerosols in the region of low cloud formation indicates that participating aerosols are produced locally. In the troposphere ionization is the main source of ultra-fine aerosol particles (<20nm) formation, and the subsequent growth into the matured aerosol distributions which act as CCN [49].

Ionization density, height profiles of the atmosphere and the major ion species produced are shown in Figure (3) [31,55]. The contributions of solar EUV and cosmic rays are clearly marked. The dominant effect of solar EUV is in the ionosphere height whereas ionization in the troposphere and stratosphere is caused by cosmic rays. Different types of ions present at various heights are also mentioned. However, the density distribution with heights of each constituent ion is not depicted as it is very complex and proper measurements are not available. *Beig and Brasseur* [61] have developed a model of troposphere ion composition based on the limited available rate coefficients and containing 20 positive ions and 29 negative ions and showed that in most atmospheric environment hyridinated clustered ions [ $H^+$ $(NH_3)_x$ (pyridine). $(H_2O)_y$ ] are the dominant positive ions. The most abundant negative ions calculated by the model are $NO_3^-$ or $HSO_4^-$ as the core, with $HNO_3$ and $H_2SO_4$ as legends. During the solar cycle, ion production varies by about 20 – 25% in the upper troposphere and ~5 – 10% in the lower troposphere at the high latitudes. The same for low latitude is ~ 4 – 7% in the upper troposphere and ~3 – 5% in the lower troposphere [62]. *Yu* [44] has studied the effect of such variation on the production of ultra-fine particles. He has simulated a size resolved multi-component aerosol system via a unified collisional mechanism involving both neutral and charged particles down to molecular sizes.

The above discussion shows that the complexity of types of ions and their occurrence increases with decreasing altitude. The ions produced within the troposphere by cosmic rays are potentially important for aerosol production in the atmosphere [31]. However, it is not simple and straightforward to show that the variability in atmospheric ionization due to the GCR flux could have a significant effect on either aerosol production or droplet growth. The growth of ion clusters is determined by the relative rates of ion-ion recombination and ion-neutral reaction. *Yu and Turco* [10,31] showed that charged molecular clusters can grow significantly faster than neutral clusters. The ultra-fine aerosol concentration is controlled by a competition between new particle formation and scavenging by the larger pre-existing background aerosol.

Simulation models have revealed that nucleation through ion-ion recombination is capable of maintaining a background aerosol distribution with realistic concentrations in the troposphere [63]. In the absence of other potential sources, we expect that a systematic change in cosmic ray ionization could lead to a change in aerosol population acting as CCN and thus influence cloud formation process. The process of aerosols acting as a source of CCN for low clouds is quite different from the source of nuclei for higher altitude ice clouds. Therefore, the influence of ionization on aerosol processes affecting low cloud CCN need not be important for high clouds [60]. For example, in the upper troposphere the short lived organic precursor gases may be less important because they may be destroyed during transport from the surface to the upper troposphere. Further, the atmospheric trace gas ammonia, which promotes sulphuric acid nucleation in the continental ground level air may not be abundantly available in the upper atmosphere due to its large solubility in liquid water and hence wet removal.

**5. Cosmic rays and lightning**

The ions generated by cosmic rays and other sources play crucial role in initiation of the cloud electrification process by convection mechanism [64]. Concentration of the natural excess of positive space charge near the ground determines the convection charging current in the initial stages of storm [65]. Similarly, the ions produced in the region above cloud top contribute to the charging current flowing with the down draft in the cloud. Further, accumulation of charges on the lower and upper layers of stratiform clouds also contributes to the charging of the cloud layer. In addition, the inductive charging theory involving the interaction of ions and precipitation particles can also contribute to the weak electrification of cloud in the initial stages. It is believed that such initial electrification processes eventually play important role in the electrification of clouds [18,66,67].

Cosmic rays produce extensive air showers by interacting with nuclei of ambient air, which may induce electrical discharge in thunderstorm [68]. *Ermakov and Stozhkov* [2] explained the lightning in terms of development of avalanche electric field and its propagation along the ionized path created by the secondary charged air shower particles. *Gurevich et al.* [69] and *Gurevich and Zybin* [70] proposed another mechanism in which the relativistic electrons are accelerated in the strong electric fields of thundercloud and produce avalanche.

## 6. Cosmic rays and stratospheric ozone

Energetic particle events including cosmic rays penetrate the terrestrial atmosphere and perturb its chemical stability. Specially, the balance of nitrogen (NOx) and hydrogen (HOx) components are changed and the $O_3$ destruction begins via catalytic process [71,72]. The effect can be synthesized as follows: (i) the ozone response to solar proton event is generally observed at high latitudes where the shielding action of the geomagnetic field is reduced, (ii) solar CRs with energies greater than 10 MeV affect the stratosphere chemistry, (iii) ozone depletion starts within few hours of the arrival of charged particles, (iv) the solar particle induced effects in the atmosphere could last days or weeks, but no relevant long-lived effects were claimed.

*Kozin et al.* [73] showed that during Forbush decreases the total ozone content, registered by 29 stations situated in the latitude range $35^0 - 60^0$, decreased practically synchronously with the galactic CR intensity. Contrary to it, *Shumilov et al.* [74] have shown an increase of the total ozone up to 10% at high latitudes and an insignificant effect at mid latitude during Forbush decreases (Fds). *Watanabe* [75] reported very small change in the stratosphere and troposphere conditions during Fds and geomagnetic activity at Syowa (69º S, 40º E). The total ozone response to major geomagnetic storms was explained in terms of changes in atmospheric dynamics [76]. The changes in circulation pattern agree with changes in total ozone. Due to limited data, it is difficult to distinguish Fds effects from those associated with the geomagnetic activity. *Marcucci et al.* [77] showed a delayed positive correlation between enhanced auroral activity and total ozone depletion over the Austral region with delay time lying between 8 to 13 days. Further, relativistic electron precipitation events occurring during geomagnetic storms, also show their imprints in the mesospheric and upper stratospheric ozone content [78]. Recently *Miroshnichenko* [7] has discussed many examples of ozone depletion associated with cosmic rays and solar proton events. *Kirillov et al.* [79] have studied the effect of energetic solar proton events on the chemical composition of the middle atmosphere during the event of December 13, 2006 and showed a good agreement between derived depletion of ozone content with experimental data obtained by the Microwave Limb Sounder instrument on the AURA spacecraft. Thus, the cosmic ray flux could be one of the potential sources for the variability in total ozone content at high latitudes and to some extent in mid-latitudes.

**7. Cosmic rays and space weather**

Space weather refers to conditions on the Sun and in the solar wind, terrestrial magnetosphere, ionosphere, and thermosphere that can influence the performance and reliability of space-borne and ground-based technological systems and can endanger human life or health [80]. There are two ways the cosmic rays affect our space weather systems: (i) the high energy charged particles produce direct changes in materials exposed to them both in space and on the surface of the Earth, (ii) the anisotropic distribution of cosmic rays incident in the Earth's atmosphere affects atmospheric processes differently which manifests in latitudinal dependence of cloud formation and hence climate change.

The space weather effects in many cases are related to geomagnetic storms. Usually a storm is said to occur when a sufficiently intense and long-lasting interplanetary convection electric field, through a substantial energization in the magnetosphere-ionsosphere system, leads to an intensified ring current strong enough to exceed some threshold value quantifying the storm "Dst" index. Cosmic rays having energy greater than 500 Mev detected by neutron monitors can be used for short-term alert systems for the major geomagnetic storms. This is based on the fact that the Larmor radii of cosmic ray particles in the interplanetary magnetic fields are sufficiently large that they are affected by large scale interplanetary magnetic field inhomogeneities originated at the Sun well before the disturbances reach the Earth's orbit and geomagnetic disturbances occur. It is also possible that the cosmic ray anisotropies may serve as a tool for remote sensing of the magnetic field changes in the heliosphere. Measurements show that CR decreases of a few percents occur at several stations at the time of geomagnetic storms [81]. However, some magnetic storms are not accompanied by decreases in CR intensity [81-84]. Therefore, more data analysis and detailed modeling work is required in this field, before CR can be used as a tool for the sensing and forecasting of geomagnetic storm, a vital aspect in space weather phenomena.

Cosmic rays and energetic charged particles in space affect the phenomena occurring in solar atmosphere, heliosphere and geosphere. These particles are relevant for (i) studying changes in the physical state of the magnetosphere and the near Earth inter-planetary medium, (ii) studying effects on space-craft and air-craft electronics, (iii) understanding of the electromagnetic propagation in the Earth environment. In fact high energy particles can deposit enough charge in the electronic components to affect the memory state (SEU : single event upset) or to induce false signals, whereas less energetic events affect components by causing permanent damage [81, 85, 86]. High energy electrons accumulate charges in poorly conductive material of the space-craft and may lead to dielectric breakdown and material damage.

## 8. Summary

In this paper we have briefly described the present understanding of atmospheric processes affected by cosmic rays. These particles produce ionization in the lower atmosphere that may influence the optical transparency of the atmosphere by changing the aerosol distribution and by acting as a source of cloud-condensation nuclei and thus affect cloud formation. This directly affects the temperature and hence climate of the atmosphere. The variation in ionization due to variation in cosmic ray activity is more prevalent in the lower atmosphere and hence modeling work should be concentrated in this region. Some important unresolved issues are :

(a) It is not known how additional ionization produced by cosmic rays could enhance charging rate and charge moment of thunderstorms.

(b) The effect of space charge on the microphysics of clouds is not well known. This can be tested by high-resolution space charge measurements from the tops to the bottoms of the clouds, together with

        measurements of particle and droplet parameters and precipitation currents.

(c)      Cosmic rays affect only low-level clouds near the equator. This needs to be understood.

(d)      Detailed investigation of the effect of ion-induced nucleation and electro- scavenging on cloud cover and precipitation is required.

(e)      The sensitivity of cloud formation to changes in background ultra-fine aerosol concentration is not known.

(f)      A realistic model of the global electric circuit including stratospheric aerosols, cloud cover, upward lightning, cosmic rays and solar wind as inputs needs to be developed.

**Acknowledgements**

Part of the work for the article was done when Devendraa Siingh was at the Center for Sun-Climate Research (CSCR), Danish National Space Institute, Copenhagen, Denmark. Devendraa Siingh thanks to Dr Jens Olaf P. Pedersen for inviting at CSCR, Denmark for a short period. We thank to Prof. B.N. Goswami, Director, IITM, Pune, India for valuable suggestions, discussions and encouragement during the preparation and revision of the manuscript. DS acknowledges the financial support from Ministry of Earth Sciences (MoES), Government of India, New Delhi.

The authors thank the anonymous reviewer for his critical comments which helped in improving the scientific value of this article.

**References**

[1] C Frohlich and J Lean, Total solar irradiance variation in new eyes to see inside the sun and stars. Deubner et al (ed), *Proc. IAU Symposium*, p. 185, Kyoto (Kluwer) 89 (1997)

[2] V I Ermakov and A V Komozokov, Charged particles measurements in the equatorial, middle and polar latitude. Trudy ZAO 179, 73 (Russian) (1992)

[3] K S Carslaw, R G Harrison and J Kirkby, *Science* **298**, 1732 (2002)

[4] J Kirkby, *Surv Geophys*, **28,** 333-375 (2007)

[5] S V Veretenenko, and M I Pudovkin, *Geomag. Aeron.* **34**, 38 (1994)

[6] M Pudovkin and S V Babushkina, *J. Atmos. Terr. Phys.* **54**, 841 (1992)

[7] L I Miroshnichenko, *J. Atmos. Solar Terr. Phys.* **70**, 450 (2008)

[8] H Svensmark, *Space Sci. Rev.* **93**, 155 (2000)

[9] J Lean, A Skumanich and O White, *Geophys. Res. Lett.* **19,** 3195 (1992)

[10] F Yu and R P Turco, *J. Geophys. Res.* 110, 4797 (2001)

[11] B A Tinsley, R P Rohrbaugh, M Hai and K V Beard, *J. Atmos. Sci.* **57**, 2118 (2000)

[12] S N Tripathi and R G Harrison, *Atmos. Res.* **62**, 57 (2002)

[13] B A Tinsley, Scavenging of condensation nuclei in clouds: dependence of sign of electroscavenging effect on droplet and CCN sizes, *in proceeding Int. Conf. on cloud and precipitation,* p.248, IAMAS, Bologna, 18-23 July, 2004 (2004).

[14] R E Dickinson, *Bull. Amer. Meteor. Soc.* **56**, 1240 (1975)

[15] M J Rycroft, S Israelsson and C Price, *J. Atmos. Solar-Terr. Phys.,* **62,** 1563 (2000)

[16] M J Rycroft and M. Fullekrug, *J. Atmos. Solar-Terr. Phys.* **66**, 1103 (2004)

[17] Devendraa Siingh, V Gopalakrishnan, R P Singh, A K Kamra, S Singh, V Pant, R Singh and A K Singh, *Atmos. Res.* **84**, 91 (2007)

[18] Devendraa Siingh, A K Singh, R P Patel, R Singh, R P Singh, B Veenadhari and M. Mukherjee, *Sur. Geophys.* **29**, 499 (2008)

[19] M. Sharma, *Earth Plan. Science Letter* **199,** 459 (2002)

[20] N J Shaviv, *Phys. Rev. Letter* **89,** 51102 (2002)

[21]     V Courtillot, Y Le Gallet, J L Mouel, F Fluteau and A Genevey, *Earth Planet. Sci. Lett*. **253,** 328 (2007)

[22]     F Arnold, *Space Sci. Rev*. **125**, 169 (2007)

[23]     H Svensmark, J O P Pedesen, N D Marsh, M B Enghoff, and U I Uggerhoj, *Proc. R. Soc.* **463,** 385 (2007)

[24]     S Eichkorn, S Wilhelm, H Aufmhoft, K H Wohlfrom and F Arnold, *Geophys Res Lett* **29**, 43 (2002)

[25]     S H Lee et al, *Science* **301**, 1886 (2003)

[26]     R Markson and C Price, *Atmos. Res.* **51,** 309 (1999)

[27]     J Hansen, R Ruedy, J Glascoe and M Sato, *J. Geophys. Res.* **104**, 30997, doi:10.1029/1999JD900835 (1999)

[28]     Y I Stozhkov, *J. Phys. G. Nuclear and Particle Physics* **29** 913 (2002)

[29]     J Lean, J Beer and R Bradley, *Geophys. Res. Lett*. **22**, 3195 (1995)

[30]     H Svensmark and E Friis-Christensen, *J. Atmos. Solar Terr. Phys.* **59**, 1225 (1997)

[31]     N Marsh and H Svensmark, *Phys. Rev. Lett.* **85,** 5004 (2000)

[32]     S C Kernthaler, R Toumi and J D Haigh, *Geophys. Res. Lett.* **27**, 853 (1999)

[33]     P Laut, *J. Atmos. Solar Terr. Phys.* **65,** 801 (2003)

[34]     B Sun and R S Bradley, *J. Geophys. Res.* **107**, 10.1029.2001JD00560(2002).

[35]     B Sun and R S Bradley, *J. Geophys. Res.* **109**, D14206 doi:10.1029/2003JD004479 (2004).

[36]     J E Kristjansson and J Kristiansen, *J. Geophys. Res.* **105**, 11851 (2000)

[37]     R G Harrision and D B Stephenson, *Proc. Roy. Soc.* **462**, 1221 (2006)

[38]     Z Y Kuang, Y B Jiang and Y L Yung, *Geophys. Res. Lett.* **25**, 1415(1998)

[39]     P D Farrar, *Climate Change* **47**, 7 (2000)

[40]     W B Rossow and Cairns B, *J. Climate* **31**, 305 (1995).

[41]     S K Solanki, *Astron. Geophys*. **43** 5 (2002)

[42]     J L Le Mouel, V Kossobokov and V Courtillot, *Earth Plant. Sci. Lett.* **232,** 273 (2005)

[43]     L E A Vieira and L A da Silva, *Geophys. Res. Lett.* **33**, L14802 doi:10.1029/2006GL026389 (2006)

[44]  F Yu, *J. Geophys. Res.* **107**, 1118 (2002).

[45]  F Yu and R P Turco, *Geophys. Res. Lett.* **27,** 883 (2000)

[46]  L Laakso, J M Makela, L Pirjola and M Kulmala, *J. Geophys. Res.* **107, D20,** 4427**,** doi: 10.1029/2002JD002140 (2002)

[47]  K Komsaare, U Hõrrak, H Tammet, Devendraa Siingh, M Vana, A Hirsikko and M Kulmala, *Proceeding of 13$^{th}$ International Conference on Atmospheric Electricity*- PS2-4 (2007)

[48]  S N Tripathi and R G Harrison, *Atmos Environ* **35**, 5817 (2001)

[49]  R P Turco, F Yu and J X Zhao, *J. Air. Waste Mang. Assoc.* **50,** 902 (2000)

[50]  R Reiter, Phenomena in Atmospheric and Environmental Electricity, elseiver, Amersterdam, (1992)

[51]  R G Harrison and K S Carslaw, *Rev. Geophys.* **41(3)**, 1012, 1-26 (2003)

[52]  S Sastry, *Nature* **438**, 746 (2005)

[53]  G A Bazilevskaya, M B Krainev and V S Makhmutov, *J. Atmos. Solar Terr. Phys.* **62,** 1577 (2000)

[54]  H V Nehar, *J. Geophys. Res.* **72** 1527 (1967)

[55]  A A Viggiano, and F Arnold, *Ion Chemistry and composition of the atmosphere, Handbook of Atmospheric electrodynamics. Vol 1 edited by H. Volland*, pp 1-26: CRC Press Boca Raton, Fl. (1995)

[56]  R Markson, *Nature* **291**, 304 (1981)

[57]  R G Harrison, *Sur. Geophys.* **25**, 441 (2004)

[58]  L J Gray, J D Haigh and R G Harrison, *A review of the influence of solar changes on the Earth's climate*, Hadley Centre Technical report no. **62**, (2005)

[59]  M J Rycroft, A Odzimek, N F Arnold, M Fullekrug, A Kulak and T Neubert, *J Atmos. Solar Terr. Phys.* **69,** 2485 (2007)

[60]  H R Pruppacher and J D Klett, *Microphysics of clouds and precipitation*, 2$^{nd}$ ed Kluwer Acad., Morwell Mass (1997)

[61]  G Beig and G Brasseur, *J. Geophys. Res.* **105**, 22671 (2000)

[62]  E P Ney, *Nature* 183, 451 (1959)

[63]  R P Turco, J X Zhao and F Yu, *Geophys. Res. Lett.* **25**, 635 (1998)

[64]  G Grenet, *Ann. Geophys*. **3**, 306 (1947)

[65]  C P R Saunders, *J. Appl. Meteor.* **32**, 642 (1993)

[66] Y Yair, *Space Sci. Rev.,* **137**, 119 (2008)

[67] C Saunders, *Space Sci. Rev.,* **137,** 335 (2008)

[68] V I Ermakov, *Proc.9$^{th}$ Int. Conf. on Atoms. Elect.* **Vol.2:** pp 485-488 (1992)

[69] A V Gurevich, K P Zubin and R A Roussel- Dupre, *Phys. Lett. A* **254**, 79 (1999)

[70] A V Gurevich and K P Zubin, *UFN*, **171**, 1177 (2001) (In russian).

[71] C H Jackman, M C Cerniglia, J E Nielsen, D J Allen, J M Zawodny, R D McPeters, A R Douglass, J E Rosenfield and R B Rood, *J. Geophys. Res.* **100** 11 641 (1995)

[72] C H Jackman, A R Douglass, R B Rood and R D McPeters, *J. Geophys. Res.* **95**, 7417 (1990)

[73] I D Kozin, I N Fedulina and B D Chakenov, *Geomagn. Aeron.* **35,** 423 (1995)

[74] O I Shumilov, E A, Kasatkina, O M Raspopov and K Henriksen, *Geomagn Aeron* 37, 24 (in Russian) (1997)

[75] T Watanabe, *Poster at the ISSI Workshop Cosmic Rays and Earth,* March 21–26: 1999, Bern, Switzerland (1999)

[76] J Laštovicka and P Mlch, *Adv. Space Res.* 20(5), **631** (1999)

[77] M F Marcucci, S Orsini, M Candidi and M Storini, *Phys. Chem. Earth.* **C24**, 141 (1999)

[78] L B Callis, R E Boughner, D N Baker, R A Mewaldt, J B Blake, R S Selesnick, J R Cummings, M Natarajan, G M Mason and J E Mazur, *Geophys. Res. Lett.* **23**, 1901 (1996)

[79] A S Kirillov, Yu V Balabin, E V Vashenyuk, Kh Fadel and L I Miroshnickenko, *Proc. 30$^{th}$ International Cosmic Rays Conference,* Merida, Yucatan, Mexico **2**, 129 (2007)

[80] H Lundstedt, ESA WPP, 1**48**, 107(1998)

[81] H V Cane, *Space Sci. Rev.* **93**, 55 (2000)

[82] K Kudela, M Storini, M Y Hoffer and A Belov, *Space Sci. Rev.* **93,** 153 (2000)

[83] K Kudela and M Storini, *J Atmos. Solar Terr. Phys.,* **67**, 907 (2005)

[84]   K Kudela, *Cosmic rays and space weather: Direct and Indirect reations, 30th International Cosmic rays Conference, Merida, Mexico* (2007).

[85]   A Hilgers and E J Daly, *ESA WPP,* **148**, 21 (1998)

[86]   K Johansson and P Dyryklev, ESA WPP, **148**, 29 (1998)

**Caption of the figure**

Fig. 1. Variation of vertical profiles of the cosmic rays ion production rate at various geomagnetic latitudes [54].

Fig.2 Flow chart of indicated that the processes linking the global atmospheric electrical circuit with global climate. Thick lines indicate established links and thin lines indicate suggested links [57,58].

Fig.3 Ionization density height profile of the Earth atmosphere and the major ion species at various altitudes [55].

Fig. 1

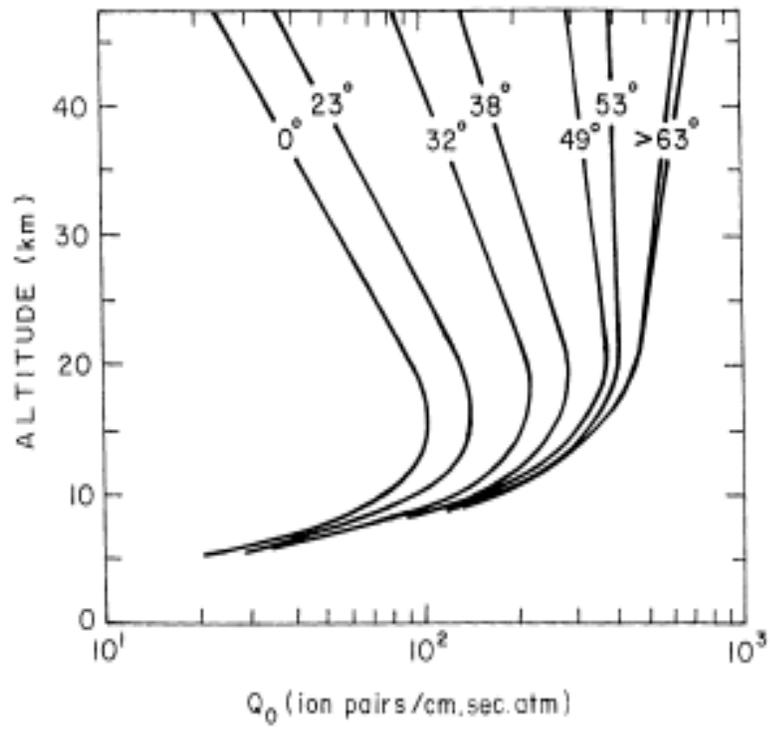

Fig.2

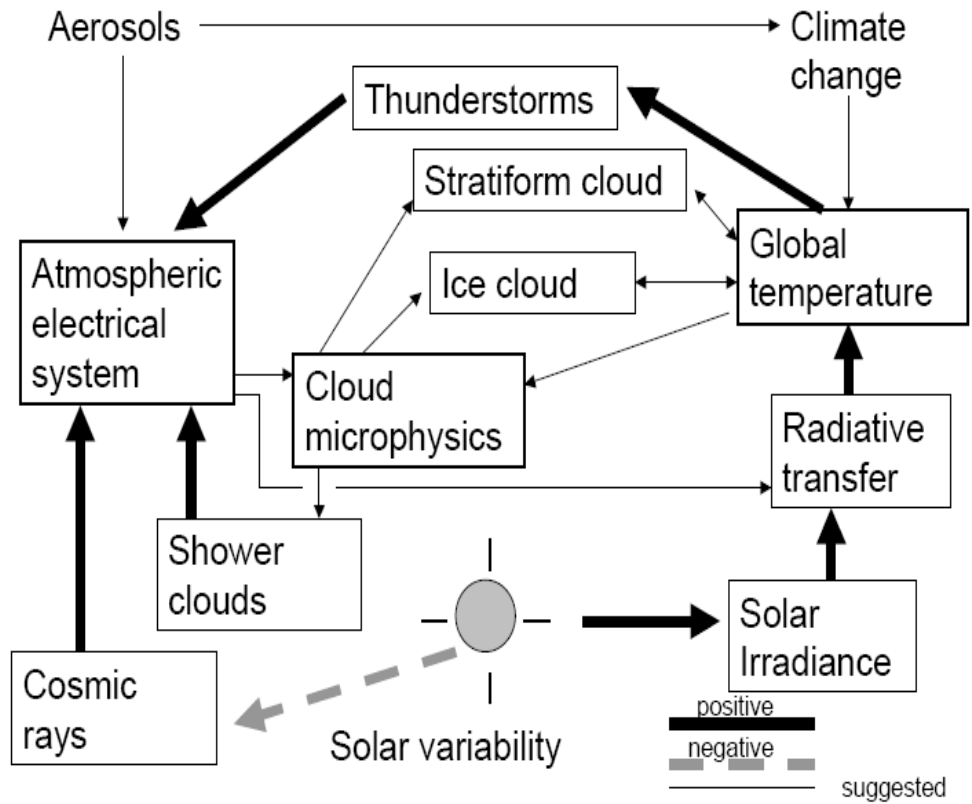

Fig.3

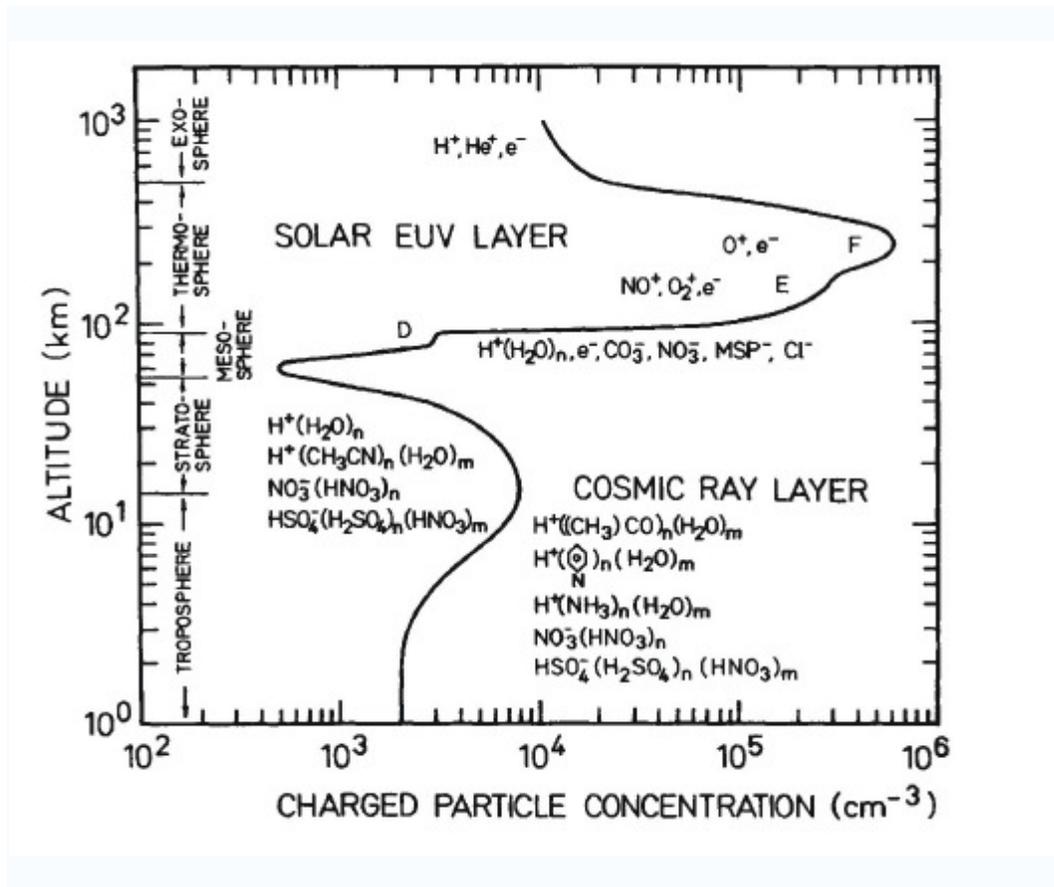